\documentclass[5p]{elsarticle}
\usepackage{amsfonts,amsmath,amssymb}
\usepackage{graphicx}
\usepackage{bm}
\usepackage[utf8]{inputenc}

\newcommand{\g}{\ensuremath{\mathfrak{g}}}

\newcommand{\B}{\ensuremath{\mathfrak{B}}}

\newcommand{\iw}{\.{I}n\"{o}n\"{u}--Wigner}

\journal{Physics Letters B}

\begin{document}

\begin{frontmatter}

\title{Standard General Relativity from Chern--Simons Gravity}

\author[ucsc]{Fernando Izaurieta}
\ead{fizaurie@ucsc.cl}

\author[udec]{Paul Minning}

\author[udec,mpi]{Alfredo P\'{e}rez}
\ead{alfperez@udec.cl}

\author[ucsc]{Eduardo Rodr\'{\i}guez}
\ead{edurodriguez@ucsc.cl}

\author[udec]{Patricio Salgado}
\ead{pasalgad@udec.cl}

% Addresses
\address[ucsc]{Departamento de Matem\'{a}tica y F\'{\i}sica Aplicadas, Universidad Cat\'{o}lica de la Sant\'{\i}sima Concepci\'{o}n, Alonso de Ribera 2850, Concepci\'{o}n, Chile}

\address[udec]{Departamento de F\'{\i}sica, Universidad de Concepci\'{o}n, Casilla 160-C, Concepci\'{o}n, Chile}

\address[mpi]{Max-Planck-Institut f\"{u}r Gravitationsphysik, Albert Einstein Institute. Am M\"{u}hlenberg~1, D-14476 Golm bei Potsdam, Germany}

\begin{abstract}
Chern--Simons models for gravity are interesting because they provide with a truly gauge-invariant action principle in the fiber-bundle sense. So far, their main drawback has largely been the perceived remoteness from standard General Relativity, based on the presence of higher powers of the curvature in the Lagrangian (except, remarkably, for three-dimensional spacetime).
Here we report on a simple model that suggests a mechanism by which standard General Relativity in five-dimensional spacetime may indeed emerge at a special critical point in the space of couplings, where additional degrees of freedom and corresponding ``anomalous'' Gauss--Bonnet constraints drop out from the Chern--Simons action.
To achieve this result, both the Lie algebra \g\ and the symmetric \g-invariant tensor that define the Chern--Simons Lagrangian are constructed by means of the Lie algebra $S$-expansion method with a suitable finite abelian semigroup $S$.
The results are generalized to arbitrary odd dimensions, and the possible extension to the case of eleven-dimensional supergravity is briefly discussed.
\end{abstract}

%\begin{keyword}



%\end{keyword}

\end{frontmatter}

\section{Introduction}

Three of the four fundamental forces of nature are consistently described by
Yang--Mills (YM) quantum theories. Gravity, the fourth fundamental interaction,
resists quantization in spite of General Relativity (GR) and YM theories
having a similar geometrical foundation. There exists, however, a very important
difference between YM theory and GR (for a thorough discussion, see, e.g., Ref.~\cite{Rov04}).

YM theories rely heavily on the existence of the ``stage'' ---the fixed, non-dynamical,
background metric structure with which the spacetime manifold $M$ is assumed to be endowed.

In GR the spacetime is a dynamical object which has independent degrees of
freedom, and is governed by dynamical equations, namely the Einstein field equations.
This means that in GR the geometry is dynamically determined.
Therefore, the construction of a gauge theory of gravity requires an action that does not consider a fixed spacetime background.

An action for gravity fulfilling the above conditions, albeit only in odd-dimensional spacetime, $d=2n+1$, was proposed long ago by Chamseddine~\cite{Cha89,Cha90} (see also Refs.~\cite{Lan38,Lov71}).
In the first-order formalism, where the independent fields are the vielbein $e^{a}$ and the spin connection $\omega^{ab}$,
the Lagrangian can be written as%
\footnote{In an effort to lighten the notation, we consistently omit the $\wedge$ symbol, but wedge product between differential forms is nevertheless assumed throughout.}
\begin{align}
L_{\text{G}}^{\left( 2n+1 \right)} & =
\kappa \varepsilon_{a_{1} \cdots a_{2n+1}} \sum_{k=0}^{n} \frac{c_{k}}{\ell^{2 \left( n-k \right) + 1}}
R^{a_{1} a_{2}} \cdots \nonumber \\
& \cdots R^{a_{2k-1} a_{2k}} e^{a_{2k+1}} \cdots e^{a_{2n+1}},
\label{eLcham}
\end{align}
where $\kappa$ and $c_{k}$ are dimensionless constants\footnote{In natural units, where $c = \hbar = 1$.} and $\ell$ is a length parameter.
As it stands, the Lagrangian~(\ref{eLcham}) is invariant under the local Lorentz transformations
\begin{align}
\delta e^{a} & = \lambda^{a}_{\phantom{a} b} e^{b}, \\
\delta \omega^{ab} & = - \mathrm{D}_{\omega} \lambda^{ab},
\end{align}
where $\lambda^{ab} = -\lambda^{ba}$ are the real, local, infinitesimal parameters that define the transformation and $\mathrm{D}_{\omega}$ stands for the Lorentz covariant derivative.

When the $c_{k}$ constants are chosen as
\begin{equation}
c_{k} = \frac{1}{2 \left( n-k \right) + 1} \binom{n}{k},
\end{equation}
then the Lagrangian~(\ref{eLcham}) can be regarded as the Chern--Simons (CS) form for the anti-de Sitter (AdS) algebra, and its invariance is accordingly enlarged to include AdS ``boosts.'' CS gravities have been extensively studied; see, for instance,
Refs.~\cite{Cha89,Cha90,Wit88,Wit89,Tro97,Tro98,Tro99,Cri00,Sal01,Sal02,Sal03a,Sal03b,Sal03c,Iza06c}.

If CS theories are to provide the appropriate gauge-theory framework
for the gravitational interaction, then these theories must satisfy the
correspondence principle, namely they must be related to GR.

An interesting research in this direction has been recently carried out~\cite{Ede06a,Has08a}.
In these references it was found that the modification of the CS theory for AdS gravity following the expansion method of Ref.~\cite{deAz02} is not sufficient to produce a direct link with GR.
In fact, it was shown that, although the action reduces to Einstein--Hilbert (EH) when the matter fields are switched off,
the field equations do not. Indeed, the corresponding field equations impose severe restrictions on the geometry, which are so strong
as to rule out, for instance, the five-dimensional Schwarzschild solution.

It is the purpose of this paper to show that standard, five-dimensional GR (without a cosmological constant) can be embedded in a CS theory for a certain Lie algebra \B. The CS Lagrangian is built from a \B-valued, one-form gauge connection $\bm{A}$ [cf.~eq.~(\ref{eAB})] which depends on a scale parameter $\ell$---a coupling constant that characterizes different regimes within the theory. The \B\ algebra, on the other hand, is constructed from the AdS algebra and a particular semigroup $S$ by means of the $S$-expansion procedure introduced in Refs.~\cite{Iza06b,Iza09a}. The field content induced by \B\ includes the vielbein $e^{a}$, the spin connection $\omega^{ab}$ and two extra bosonic fields $h^{a}$ and $k^{ab}$. The full CS field equations impose severe restrictions on the geometry~\cite{Ede06a,Has08a},                                      which at a special critical point in the space of couplings ($\ell = 0$) disappear to yield pure GR.

The paper is organized as follows.
In Sec.~\ref{scsadsg} we briefly review CS AdS gravity.
An explicit action for five-dimen\-sional gravity is considered in Sec.~\ref{s5d},
where the Lie algebra $S$-expansion procedure is used to obtain a \B-invariant CS action that includes the coupling constant $\ell$.
It is then shown that the usual EH theory arises in the strict limit where the scale parameter $\ell$ equals zero.
Sec.~\ref{sfin} concludes the work with a comment about possible developments.

\section{\label{scsadsg}Chern--Simons anti-de Sitter Gravity}

The CS AdS Lagrangian for gravity in $d=2n+1$ dimensions is given by~\cite{Cha89,Cha90}
\begin{align}
L_{\text{AdS}}^{\left( 2n+1 \right)} & =
\kappa \varepsilon_{a_{1} \cdots a_{2n+1}} \sum_{k=0}^{n} \frac{c_{k}}{\ell^{2 \left( n-k \right) + 1}}
R^{a_{1} a_{2}} \cdots \nonumber \\
& \cdots R^{a_{2k-1} a_{2k}} e^{a_{2k+1}} \cdots e^{a_{2n+1}}
\label{AdSGrav}
\end{align}
where the $c_{k}$ constants are defined as
\begin{equation}
c_{k} = \frac{1}{2 \left( n-k \right) + 1} \binom{n}{k},
\end{equation}
$e^{a}$ corresponds to the one-form \emph{vielbein}, and
$R^{ab} = \mathrm{d} \omega ^{ab} + \omega _{\phantom{a}c}^{a} \omega^{cb}$
to the Riemann curvature in the first-order formalism.

The Lagrangian~(\ref{AdSGrav}) is off-shell invariant under the AdS Lie algebra
$\mathfrak{so} \left( 2n,2 \right)$,
whose generators $\tilde{\bm{J}}_{ab}$ of
Lorentz transformations and $\tilde{\bm{P}}_{a}$ of AdS boosts
satisfy the commutation relations
\begin{align}
\left[ \tilde{\bm{J}}_{ab}, \tilde{\bm{J}}_{cd} \right] & =
\eta_{cb} \tilde{\bm{J}}_{ad} - \eta_{ca} \tilde{\bm{J}}_{bd} +
\eta_{db} \tilde{\bm{J}}_{ca} - \eta_{da} \tilde{\bm{J}}_{cb}, \\
\left[ \tilde{\bm{J}}_{ab}, \tilde{\bm{P}}_{c} \right] & =
\eta_{cb} \tilde{\bm{P}}_{a} - \eta_{ca} \tilde{\bm{P}}_{b}, \\
\left[ \tilde{\bm{P}}_{a}, \tilde{\bm{P}}_{b} \right] & =
\tilde{\bm{J}}_{ab}.
\end{align}

The Levi-Civita symbol $\varepsilon_{a_{1} \cdots a_{2n+1}}$ in~(\ref{AdSGrav})
is to be regarded as the only non-vanishing component of the symmetric,
$\mathfrak{so} \left( 2n,2 \right)$-invariant tensor of rank $r=n+1$, namely
\begin{equation}
\left\langle \tilde{\bm{J}}_{a_{1}a_{2}} \cdots \tilde{\bm{J}}_{a_{2n-1}a_{2n}} \tilde{\bm{P}}_{a_{2n+1}} \right\rangle =
\frac{2^{n}}{n+1} \varepsilon_{a_{1} \cdots a_{2n+1}}.
\end{equation}

In order to interpret the gauge field associated with a translational
generator $\tilde{\bm{P}}_{a}$ as the vielbein, one is \emph{forced}
to introduce a length scale $\ell$ in the theory. To see why this happens,
consider the following argument. Given that (i)~the exterior derivative operator
$\mathrm{d} = \mathrm{d} x^{\mu} \partial_{\mu}$
is dimensionless, and (ii)~one can always choose Lie algebra generators $\bm{T}_{A}$ to be dimensionless as well,
the one-form connection fields
$A^{A} = A_{\phantom{A}\mu}^{A} \mathrm{d} x^{\mu}$
must also be dimensionless.
However, the vielbein
$e^{a} = e_{\phantom{a}\mu}^{a} \mathrm{d} x^{\mu}$ must have dimensions of length if it is
to be related to the spacetime metric $g_{\mu \nu}$ through the usual equation
$g_{\mu \nu} = e_{\phantom{a}\mu}^{a} e_{\phantom{b}\mu}^{b} \eta_{ab}$.
This means that the ``true'' gauge
field must be of the form $e^{a}/\ell$, where $\ell$ is a length.

Therefore, following Refs.~\cite{Cha89,Cha90}, the one-form gauge field $\bm{A}$ of the CS theory is given in this case by
\begin{equation}
\bm{A} = \frac{1}{\ell} e^{a} \tilde{\bm{P}}_{a} +
\frac{1}{2} \omega^{ab} \tilde{\bm{J}}_{ab}.
\label{ehcs0}
\end{equation}

It is important to notice that once the length scale $\ell$ is brought in to the CS theory,
the Lagrangian splits into several sectors, each one of them proportional to a different power of $\ell$,
as we can see directly in eq.~(\ref{AdSGrav}).

CS gravity is a well-defined gauge theory, but the presence of
higher powers of the curvature makes its dynamics very remote from that for
standard EH gravity. In fact, it seems very difficult to recover
EH dynamics from a pure gauge, \emph{off-shell} invariant theory in odd%
\footnote{In even dimensions, the problem has been solved in a very elegant way using
topological defects~\cite{Ana07}.}
dimensions (see, e.g., Refs.~\cite{Ede06a,Has08a}).

\section{\label{s5d}Einstein--Hilbert Action from five-dimensional Chern--Simons gravity}

In this section we show how to recover five-dimensional GR from CS Gravity.
The generalization to an arbitrary odd dimension is given in Appendix~\ref{appex}.

\subsection{$S$-Expansion Procedure}

The Lagrangian for five-dimensional CS AdS gravity can be written as
\begin{align}
L_{\text{AdS}}^{\left( 5 \right)} & =
\kappa \varepsilon_{abcde} \left(
\frac{1}{5\ell^{5}} e^{a} e^{b} e^{c} e^{d} e^{e} +
\frac{2}{3\ell^{3}} R^{ab} e^{c} e^{d} e^{e} +
\right. \nonumber \\ & \left.
+ \frac{1}{\ell} R^{ab} R^{cd} e^{e}
\right).
\label{ehcs4}
\end{align}
From this Lagrangian it is apparent that neither the $\ell \rightarrow \infty$
nor the $\ell \rightarrow 0$ limits yield the EH term
$\varepsilon_{abcde} R^{ab} e^{c} e^{d} e^{e}$ alone.
Rescaling $\kappa$ properly, those limits will lead to either the
Gauss--Bonnet term (Poincar\'{e} CS gravity) or the cosmological
constant term by itself, respectively.

The Lagrangian~(\ref{ehcs4}) is arrived at as the CS form for the AdS algebra in five dimensions.
This algebra choice is crucial, since it permits the interpretation of the gauge fields $e^{a}$ and $\omega^{ab}$
as the f\"{u}nfbein and the spin connection, respectively. It is, however, not the only possible choice:
as we explicitly show below, there exist other Lie algebras that also allow for a similar identification and lead to a CS Lagrangian that touches upon EH in a certain limit.

Following the definitions of Ref.~\cite{Iza06b}, let us consider the $S$-expansion of the Lie algebra
$\mathfrak{so} \left( 4,2 \right)$ using $S_{\mathrm{E}}^{\left( 3 \right) }$ as the relevant finite abelian semigroup.
After extracting a resonant subalgebra and performing its $0_{S}$-reduction,
one finds a new Lie algebra, call it \B, with the desired properties.
In simpler terms, consider the Lie algebra generated by
$\left\{ \bm{J}_{ab}, \bm{P}_{a}, \bm{Z}_{ab},\bm{Z}_{a} \right\}$,
where these new generators can be written as
\begin{align}
\bm{J}_{ab} & = \lambda_{0} \otimes \tilde{\bm{J}}_{ab}, \\
\bm{Z}_{ab} & = \lambda_{2} \otimes \tilde{\bm{J}}_{ab}, \\
\bm{P}_{a}  & = \lambda_{1} \otimes \tilde{\bm{P}}_{a},  \\
\bm{Z}_{a}  & = \lambda_{3} \otimes \tilde{\bm{P}}_{a}.
\end{align}
Here $\tilde{\bm{J}}_{ab}$ and $\tilde{\bm{P}}_{a}$
correspond to the original generators of $\mathfrak{so} \left( 4,2 \right)$,
and the $\lambda_{\alpha}$ belong to a finite abelian semigroup.
The semigroup elements $\left\{ \lambda _{0}, \lambda _{1}, \lambda _{2}, \lambda _{3}, \lambda _{4} \right\}$
are \emph{not} real numbers and they are \emph{dimensionless}.
In this particular case, they obey the multiplication law
\begin{equation}
\lambda_{\alpha} \lambda_{\beta} = \left\{
\begin{array}{ll}
\lambda_{\alpha + \beta}, & \text{when } \alpha + \beta \leq 4, \\
\lambda_{4}, & \text{when } \alpha + \beta >4.
\end{array}
\right.
\end{equation}

An explicit matrix representation for the $\lambda_{\alpha}$ is given in Table~\ref{tmats}.

\begin{table}
\caption{Explicit matrix representation for the finite abelian semigroup $S_{\mathrm{E}}^{\left( 3 \right) }$.}
\begin{align}
\lambda_{0} & = \left(
\begin{array}{cccc}
1 & 0 & 0 & 0 \\
0 & 1 & 0 & 0 \\
0 & 0 & 1 & 0 \\
0 & 0 & 0 & 1
\end{array}
\right), \\
\lambda_{1} & = \left(
\begin{array}{cccc}
0 & 0 & 0 & 0 \\
1 & 0 & 0 & 0 \\
0 & 1 & 0 & 0 \\
0 & 0 & 1 & 0
\end{array}
\right), \\
\lambda_{2} & = \left(
\begin{array}{cccc}
0 & 0 & 0 & 0 \\
0 & 0 & 0 & 0 \\
1 & 0 & 0 & 0 \\
0 & 1 & 0 & 0
\end{array}
\right), \\
\lambda_{3} & = \left(
\begin{array}{cccc}
0 & 0 & 0 & 0 \\
0 & 0 & 0 & 0 \\
0 & 0 & 0 & 0 \\
1 & 0 & 0 & 0
\end{array}
\right), \\
\lambda_{4} & = \left(
\begin{array}{cccc}
0 & 0 & 0 & 0 \\
0 & 0 & 0 & 0 \\
0 & 0 & 0 & 0 \\
0 & 0 & 0 & 0
\end{array}
\right).
\end{align}
\label{tmats}
\end{table}

Using Theorem~VII.2 from Ref.~\cite{Iza06b}, it is possible to show that the
only non-vanishing components of a symmetric invariant tensor for the \B\ algebra
are given by
\begin{align}
\left\langle \bm{J}_{ab} \bm{J}_{cd} \bm{P}_{e} \right\rangle & =
\frac{4}{3} \ell^{3} \alpha_{1} \varepsilon_{abcde}, \\
\left\langle \bm{J}_{ab} \bm{J}_{cd} \bm{Z}_{e} \right\rangle & =
\frac{4}{3} \ell^{3} \alpha_{3} \varepsilon_{abcde}, \\
\left\langle \bm{J}_{ab} \bm{Z}_{cd} \bm{P}_{e} \right\rangle & =
\frac{4}{3} \ell^{3} \alpha_{3} \varepsilon_{abcde},
\end{align}
where $\alpha_{1}$ and $\alpha_{3}$ are arbitrary independent constants of dimension $\left[ \text{length} \right]^{-3}$.

In order to write down a CS Lagrangian for the \B\ algebra, we start from the \B-valued, one-form gauge connection
\begin{equation}
\bm{A} = \frac{1}{2} \omega^{ab} \bm{J}_{ab} +
\frac{1}{\ell} e^{a} \bm{P}_{a} +
\frac{1}{2} k^{ab} \bm{Z}_{ab} +
\frac{1}{\ell} h^{a} \bm{Z}_{a},
\label{eAB}
\end{equation}
and the associated two-form curvature
\begin{align}
\bm{F} & = \frac{1}{2} R^{ab} \bm{J}_{ab} +
\frac{1}{\ell} T^{a} \bm{P}_{a} +
\frac{1}{2} \left( \mathrm{D}_{\omega} k^{ab} + \frac{1}{\ell^{2}} e^{a} e^{b} \right) \bm{Z}_{ab} +
\nonumber \\ &
+ \frac{1}{\ell} \left( \mathrm{D}_{\omega} h^{a} + k_{\phantom{a}b}^{a} e^{b} \right) \bm{Z}_{a}.
\end{align}

Consistency with the dual procedure of $S$-expansion in terms of the
Maurer--Cartan (MC) forms~\cite{Iza09a} demands that $h^{a}$ inherits units
of length from the f\"{u}nfbein; this is why it is necessary to
introduce the $\ell$ parameter again, this time associated to $h^{a}$.

It is interesting to observe that $\bm{J}_{ab}$ are still Lorentz
generators, but $\bm{P}_{a}$ are no longer AdS boosts; in fact, we have
$\left[ \bm{P}_{a}, \bm{P}_{b} \right] = \bm{Z}_{ab}$.
However, $e^{a}$ still transforms as a vector under Lorentz transformations, as it must be
in order to recover gravity in this scheme.

\subsection{The Lagrangian}

Using the extended Cartan homotopy formula as in Ref.~\cite{Iza06a}, and
integrating by parts, it is possible to write down the CS
Lagrangian in five dimensions for the \B\ algebra as
\begin{align}
L_{\text{CS}}^{\left( 5 \right)} & = \alpha _{1} \ell^{2} \varepsilon_{abcde} R^{ab} R^{cd} e^{e} +
\alpha_{3} \varepsilon_{abcde} \left( \frac{2}{3} R^{ab} e^{c} e^{d} e^{e} +
\right. \nonumber \\ & \left. +
2 \ell^{2} k^{ab} R^{cd} T^{e} +
\ell^{2} R^{ab} R^{cd} h^{e} \right).
\label{ehcs21'}
\end{align}
Two important points can now be made:
\begin{enumerate}
\item The Lagrangian~(\ref{ehcs21'}) is split in two independent pieces, one
proportional to $\alpha_{1}$ and the other proportional to $\alpha_{3}$.
The piece proportional to $\alpha_{1}$ corresponds to the \iw\ contraction of the Lagrangian~(\ref{ehcs4}),
and therefore it is the CS Lagrangian for the Poincar\'{e} Lie group $\text{ISO} \left( 4,1 \right)$.
The piece proportional to $\alpha_{3}$ contains the EH term $\varepsilon_{abcde} R^{ab} e^{c} e^{d} e^{e}$
plus non-linear couplings between the curvature and the bosonic ``matter'' fields $k^{ab}$ and $h^{a}$.
These couplings are all proportional to $\ell^{2}$.

\item When the constant $\alpha_{1}$ vanishes, the Lagrangian~(\ref{ehcs21'}) almost
exactly matches the one given in Ref.~\cite{Ede06a}, the only difference being
that in our case the coupling constant $\ell^{2}$ appears explicitly in the last two terms.
This difference has its origin in the fact that, in Ref.~\cite{Ede06a}, both the
symmetry and the Lagrangian arise through the process of Lie algebra \emph{expansion}
(see Ref.~\cite{deAz02}), using $1/\ell$ as an expansion \emph{parameter}.
In contrast, no parameter has been used here to create the
new \B-symmetry and the Lagrangian. Instead, they were constructed through the
\emph{$S$-expansion} procedure, using the \emph{dimensionless}
elements of a \emph{finite abelian semigroup} (which in general cannot
be represented by real numbers, but rather by matrices).
\end{enumerate}

The presence or absence of the coupling constant $\ell$ in the Lagrangian~(\ref{ehcs21'}) may seem like a minor or trivial
matter, but it is not. As the authors of Ref.~\cite{Ede06a} clearly
state, the presence of the EH term in this kind of action
does not guarantee that the dynamics will be that of GR.
In general, extra constraints on the geometry appear, even around a
``vacuum'' solution with $k^{ab} = 0$, $h^{a} = 0$. In fact, the variation of the
Lagrangian, modulo boundary terms, can be written as
\begin{align*}
\delta L_{\text{CS}}^{\left( 5 \right)} & =
\varepsilon_{abcde} \left( 2 \alpha_{3} R^{ab} e^{c} e^{d} +
\alpha_{1} \ell^{2} R^{ab} R^{cd} + \right. \\
& \left. + 2 \alpha_{3} \ell^{2} \mathrm{D}_{\omega} k^{ab} R^{cd} \right) \delta e^{e}
+ \alpha_{3} \ell^{2} \varepsilon_{abcde} R^{ab} R^{cd} \delta h^{e} + \\
& + 2 \varepsilon_{abcde} \delta \omega^{ab} \left( \alpha_{1} \ell^{2} R^{cd} T^{e} +
\alpha_{3} \ell^{2} \mathrm{D}_{\omega} k^{cd} T^{e} + \right. \nonumber \\
& \left. + \alpha_{3} e^{c} e^{d} T^{e} +
\alpha_{3} \ell^{2} R^{cd} \mathrm{D}_{\omega} h^{e} +
\alpha_{3} \ell^{2} R^{cd} k_{\phantom{e}f}^{e} e^{f} \right) + \\
& + 2 \alpha_{3} \ell^{2} \varepsilon_{abcde} \delta k^{ab} R^{cd} T^{e}.
\end{align*}

Therefore, when $\alpha_{1}$ vanishes,
the torsionless condition is imposed,
and a solution without matter ($k^{ab} = 0$, $h^{a} = 0$) is singled out,
we are left with
\begin{equation}
\delta L_{\text{CS}}^{\left( 5 \right)} = 2 \alpha_{3} \varepsilon_{abcde} R^{ab} e^{c} e^{d} \delta e^{e} +
\alpha_{3} \ell^{2} \varepsilon_{abcde} R^{ab} R^{cd} \delta h^{e}.
\label{eeom}
\end{equation}
In this way, besides the GR equations of motion
[first term in~(\ref{eeom})],
the equations of motion of pure Gauss--Bonnet theory
[second term in~(\ref{eeom})]
also in general appear as an anomalous constraint on the geometry.

It is at this point where the presence of the $\ell$ parameter makes the difference.
In the present approach, it plays the r\^{o}le of a coupling constant between geometry and ``matter.''
Remarkably, in the strict limit where the coupling constant $\ell$ equals zero,
we obtain solely the EH term in the Lagrangian
\begin{equation}
L_{\text{CS}}^{\left( 5 \right)} = \frac{2}{3} \alpha_{3} \varepsilon_{abcde} R^{ab} e^{c} e^{d} e^{e}.
\end{equation}
In the same way, in the limit where $\ell = 0$ the extra constraints just vanish,
and $\delta L_{\mathrm{CS}}^{\left( 5\right) }=0$ leads us to just
the EH dynamics in vaccum,
\begin{equation*}
\delta L_{\mathrm{CS}}^{\left( 5\right) }=2\alpha _{3}\epsilon
_{abcde}R^{ab}e^{c}e^{d}\delta e^{e}+2\alpha _{3}\epsilon _{abcde}\delta
\omega ^{ab}e^{c}e^{d}T^{e}.
\end{equation*}

It is interesting to observe that the argument given here is not just a five-dimensional accident.
In every odd dimension, it is possible to perform the $S$-expansion in the way sketched here,
take the vanishing coupling constant limit $\ell = 0$ and recover EH gravity (see Appendix~\ref{appex}).

\section{\label{sfin}Comments and Possible Developments}

The present work shows the difference between the possibilities of the $S$-expansion procedure \cite{Iza06b,Iza09a} (using semigroups)
and the MC forms expansion (using a parameter).

The $S$-expansion procedure allows us to study in a deeper way the r\^{o}le of the $\ell$ parameter.
In fact, it makes possible to recover odd-dimensional EH gravity from a CS theory in the strict limit where the coupling constant $\ell$ equals zero while keeping the effective Newton's constant fixed. It is only at this point ($\ell = 0$) in the space of couplings that the ``anomalous'' Gauss--Bonnet constraints disappear from the on-shell system.

This is in strong contrast with the standard CS AdS gravity~\cite{Cha89,Cha90} or the result of expansion using a real parameter~\cite{Ede06a,Has08a}.

The system of extra constraints on the geometry arises for any finite value of the scale parameter (coupling constant $\ell \neq 0$).
In other words, for $\ell \neq 0$ the system has to obey Einstein's equations plus a set of on-shell Gauss--Bonnet constraints.
% In the particular case of five dimensions and for small scale parameter we obtain the Einstein tensor plus small corrections
% including extra degrees of freedom equals to zero and $\epsilon R R = 0$ plus further conditions on the extra degree of freedom.
In this way, GR corresponds to a special critical point, $\ell = 0$, in the space of couplings of the CS gravitational theory.

The simple model and procedure considered here could play an important r\^{o}le
in the context of supergravity in higher dimensions. In fact, it seems
likely that it might be possible to recover the standard eleven-dimensional Crem\-mer--Julia--Scherk
Supergravity from a CS/transgression form principle, in a way
reminiscent to the one shown here. In this way, the procedure sketched here
could provide us with valuable information on what the underlying geometric
structure of Supergravity and M~theory in $d=11$ could be.

\section*{Acknowledgements}

The authors wish to thank R.~Caroca, J.~Cris\'{o}stomo and N.~Merino for enlightening discussions.
F.~I. wishes to thank J.~A.~de~Azc\'{a}rraga for his warm hospitality at the Universitat de Val\`{e}ncia, where part of this work was done, and for enlightening discussions.
One of the authors (A.~P.) wishes to thank S.~Theisen for his kind hospitality at the MPI f\"{u}r Gravitationsphysik in Golm, where part of this work was done. He is also grateful to the Deutscher Akademischer Austauschdienst (DAAD), Germany, and the Comisi\'{o}n Nacional de Investigaci\'{o}n Cient\'{\i}fica y Tecnol\'{o}gica (CONICYT), Chile, for financial support.
P.~S. was supported by Fondo Nacional de Desarrollo Cient\'{\i}fico y Tecnol\'{o}gico (FONDECYT, Chile) Grants 1080530 and 1070306 and by the Universidad de Concepci\'{o}n, Chile, through DIUC Grants 208.011.048 - 1.0.
F.~I. was supported by FONDECYT Grant 11080200, the Vicerrector\'{\i}a de Asuntos Internacionales y Cooperaci\'{o} of the Universitat de Val\`{e}ncia, Spain, and the Direcci\'{o}n de Perfeccionamiento y Postgrado of the Universidad Cat\'{o}lica de la Sant\'{\i}sima Concepci\'{o}n, Chile.
E.~R. was supported by FONDECYT Grant 11080156.

\appendix

\section{\label{appex}Extension to Higher Odd Dimensions}

The CS AdS Lagrangian for gravity in $d=2n+1$ dimensions is given by [cf.~eq.~(\ref{AdSGrav})]
\begin{align}
L_{\text{AdS}}^{\left( 2n+1 \right)} & =
\kappa \varepsilon_{a_{1} \cdots a_{2n+1}} \sum_{k=0}^{n} \frac{c_{k}}{\ell^{2 \left( n-k \right) + 1}} R^{a_{1} a_{2}}
\cdots \nonumber \\ & \cdots
R^{a_{2k-1} a_{2k}} e^{a_{2k+1}} \cdots e^{a_{2n+1}},
\label{AdSGrav2}
\end{align}
where the $c_{k}$ constants are defined as
\begin{equation}
c_{k} = \frac{1}{2 \left( n-k \right) + 1} \binom{n}{k},
\end{equation}
$e^{a}$ corresponds to the one-form vielbein, and
$R^{ab} = \mathrm{d} \omega ^{ab} + \omega _{\phantom{a}c}^{a} \omega^{cb}$
to the Riemann curvature in the first-order formalism.

Simple inspection of~(\ref{AdSGrav2}) shows that neither the $\ell \rightarrow \infty$ nor the $\ell \rightarrow 0$ limits produce EH gravity.

Let us instead consider the $S$-expansion~\cite{Iza06b} of the AdS algebra
$\mathfrak{so} \left( 2n,2 \right)$
through the abelian semigroup $S = \left\{ \lambda_{\alpha} \right\}$ defined by the product
\begin{equation*}
\lambda _{\alpha }\lambda _{\beta }=\left\{ 
\begin{array}{ll}
\lambda _{\alpha +\beta }, & \text{ when }\alpha +\beta \leq 2n, \\ 
\lambda _{2n}, & \text{ when }\alpha +\beta >2n.%
\end{array}%
\right.
\end{equation*}

The $\lambda_{\alpha}$ elements are dimensionless, and can be represented by the set of
$2n \times 2n$ sparse matrices
$\left[ \lambda_{\alpha} \right]_{\phantom{i}j}^{i} = \delta_{j+\alpha}^{i}$,
where
$i,j = 1, \ldots 2n-1$,
$\alpha = 0, \ldots 2n$,
and $\delta$ stands for the Kronecker delta.

The generators of the new Lie algebra $\B_{2n+1}$ obtained through $S$-expansion, resonant subalgebra extraction and $0_{S}$-reduction~\cite{Iza06b} can be thought of as the direct products
\begin{align}
\bm{J}_{\left( ab, 2k  \right)} & = \lambda_{2k}  \otimes \tilde{\bm{J}}_{ab}, \\
\bm{P}_{\left( a, 2k+1 \right)} & =\lambda_{2k+1} \otimes \tilde{\bm{P}}_{a},
\end{align}%
with $k=0,\ldots ,n-1.$ According to theorem~VII.2 from Ref.~\cite{Iza06b}, the
symmetric invariant tensor of order $n+1$ for this case can be chosen to be
\begin{align}
& \left\langle \bm{J}_{\left( a_{1} a_{2}, i_{1} \right)} \cdots
\bm{J}_{\left( a_{2n-1} a_{2n}, i_{n} \right)}
\bm{P}_{\left( a_{2n+1}, i_{n+1} \right)} \right\rangle \nonumber \\
& = \frac{2^{n} \ell^{2n-1}}{n+1} \alpha_{j} \delta_{i_{1} + \cdots + i_{n+1}}^{j}
\varepsilon_{a_{1} \cdots a_{2n+1}},
\end{align}
where $i_{p},j=0,\ldots ,2n-1$, the $\alpha_{i}$ are arbitrary constants, and all other components vanish.

The $\B_{2n+1}$-valued, one-form gauge connection $\bm{A}$ takes the form
\begin{equation}
\bm{A} = \sum_{k=0}^{n-1} \left[ \frac{1}{2} \omega^{\left( ab, 2k \right)} \bm{J}_{\left( ab, 2k \right)} +
\frac{1}{\ell} e^{\left( a, 2k+1 \right)} \bm{P}_{\left( a, 2k+1 \right)} \right].
\end{equation}

Using the matrix representation given above for the semigroup elements, it is possible
to show that the two-form curvature
$\bm{F} = \mathrm{d} \bm{A} + \bm{A}^{2}$
is given by
\begin{equation}
\bm{F} = \sum_{k=0}^{n-1} \left[ \frac{1}{2} F^{\left( ab, 2k \right)} \bm{J}_{\left( ab, 2k \right)} +
\frac{1}{\ell} F^{\left( a, 2k+1 \right)} \bm{P}_{\left( a, 2k+1 \right)} \right],
\end{equation}
where
\begin{align}
F^{\left( ab, 2k \right)} & = \mathrm{d} \omega^{\left( ab, 2k \right)} +
\eta_{cd} \omega^{\left( ac, 2i \right)} \omega^{\left( db, 2j \right)} \delta_{i+j}^{k} + \nonumber \\
& + \frac{1}{\ell^{2}} e^{\left( a, 2i+1 \right)} e^{\left( b, 2j+1 \right)} \delta_{i+j+1}^{k}, \\
F^{\left( a, 2k+1 \right)} & = \mathrm{d} e^{\left( a, 2k+1 \right)} +
\eta_{bc} \omega^{\left( ab, 2i \right)} e^{\left( c, 2j \right)} \delta_{i+j}^{k}.
\end{align}

Following the method presented in Ref.~\cite{Iza06a}, it is possible to write down
the CS $\B_{2n+1}$ Lagrangian explicitly as
\begin{align}
L_{\text{CS}}^{\left( 2n+1 \right)} & =
\sum_{k=1}^{n} \ell^{2k-2} c_{k} \alpha_{j}
\delta_{i_{1} + \cdots + i_{n+1}}^{j}
\delta_{p_{1} + q_{1}}^{i_{k+1}} \cdots
\delta_{p_{n-k} + q_{n-k}}^{i_{n}}
\nonumber \\ &
\varepsilon_{a_{1} \cdots a_{2n+1}}
F^{\left( a_{1} a_{2}, i_{1} \right)} \cdots
F^{\left( a_{2k-1} a_{2k}, i_{k} \right)}
e^{\left( a_{2k+1}, p_{1} \right)}
\nonumber \\ &
e^{\left( a_{2k+2}, q_{1} \right)} \cdots
e^{\left( a_{2n-1}, p_{n-k} \right)}
e^{\left( a_{2n}, q_{n-k} \right)}
e^{\left( a_{2n+1}, i_{n+1} \right)}.
\label{emons}
\end{align}

In the vanishing coupling constant limit $\ell = 0$, the only surviving term in~(\ref{emons}) is given by $k=1$:
\begin{align*}
\left. L_{\text{CS}}^{\left( 2n+1 \right)} \right\vert_{\ell = 0} & =
c_{1} \alpha_{j} \delta_{i + k_{1} \cdots + k_{2n-1}}^{j}
\varepsilon_{a_{1} \cdots a_{2n+1}} \\
& F^{\left( a_{1} a_{2}, i \right)}
e^{\left( a_{3}, k_{1} \right)} \cdots
e^{\left( a_{2n+1}, k_{2n-1} \right)} \\
& = c_{1} \alpha_{j} \delta_{2p + 2q_{1} + 1 + \cdots + 2_{q_{2n-1}} + 1}^{j}
\varepsilon_{a_{1} \cdots a_{2n+1}} \\
& F^{\left( a_{1} a_{2}, 2p \right)}
e^{\left( a_{3}, 2q_{1} + 1 \right)} \cdots
e^{\left( a_{2n+1}, 2_{q_{2n-1}} + 1 \right)} \\
& = c_{1} \alpha_{j} \delta_{2\left( p + q_{1} + \cdots + q_{2n-1} \right) + 2n-1}^{j}
\varepsilon_{a_{1} \cdots a_{2n+1}} \\
& F^{\left( a_{1} a_{2}, 2p \right)}
e^{\left( a_{3}, 2q_{1} + 1 \right)} \cdots e^{\left( a_{2n+1}, 2_{q_{2n-1}} + 1 \right)}.
\end{align*}

The only non-vanishing component of this expression occurs for
$p = q_{1} = \cdots = q_{2n-1} = 0$
and is proportional to the EH Lagrangian,
\begin{align}
\left. L_{\text{CS}}^{\left( 2n+1 \right)} \right\vert_{\ell = 0} & =
c_{1} \alpha_{2n-1} \varepsilon_{a_{1} \cdots a_{2n+1}}
F^{\left( a_{1} a_{2}, 0 \right)}
\nonumber \\ &
e^{\left( a_{3}, 1 \right)} \cdots
e^{\left( a_{2n+1}, 1 \right)} \\
& = \frac{n\alpha_{2n-1}}{2n-1} \varepsilon_{a_{1} \cdots a_{2n+1}}
R^{a_{1} a_{2}} e^{a_{3}} \cdots e^{a_{2n+1}}.
\end{align}

\end{document}